\documentclass[twocolumn,twoside,slac_two]{revtex4}
\usepackage{graphicx}
\usepackage{fancyhdr}
\newcommand{\Fermi}{\textit{Fermi}}
\newcommand{\PV}{\texttt{P6V11} }
\newcommand{\kpc}{\,\text{kpc}}

\newcommand{\s}{\,\text{s}}

\newcommand{\cm}{\,\text{cm}}
\newcommand{\km}{\,\text{km}}

\begin{document}

\title{The GeV Excess Shining Through: Background Systematics for the Inner Galaxy Analysis}


\author{F. Calore\footnote{Speaker. E-mail: f.calore@uva.nl}, C. Weniger}
\affiliation{GRAPPA Institute, University of Amsterdam, Science Park 904, 1090 GL Amsterdam, The Netherlands}
\author{I. Cholis}
\affiliation{Center for Particle Astrophysics, Fermi National Accelerator Laboratory, Batavia, IL, 60510, USA}

\begin{abstract}
Recently, a spatially extended excess of gamma rays collected by the \Fermi-LAT from the inner region of the Milky Way has been detected by different groups and with increasingly sophisticated techniques. Yet, any final conclusion about the morphology and spectral properties of such an extended diffuse emission are subject  to a number of potentially critical uncertainties, related to the high density of cosmic rays, gas, magnetic fields and abundance of point sources.
We will present a thorough study of the systematic uncertainties related to the modelling of diffuse background and to the propagation of cosmic rays in the inner part of our Galaxy. We will test a large set of models for the Galactic diffuse emission, generated by varying the propagation parameters within extreme conditions. By using those models in the fit of \Fermi-LAT data as Galactic foreground, we will show that the gamma-ray excess survives and we will quantify the uncertainties on the excess emission morphology and energy spectrum. 
\end{abstract}

\maketitle

\thispagestyle{fancy}

\section{Introduction}
\label{sec:intro}
\begin{figure*}[t!]
    \begin{center}
        \includegraphics[width=0.4\linewidth]{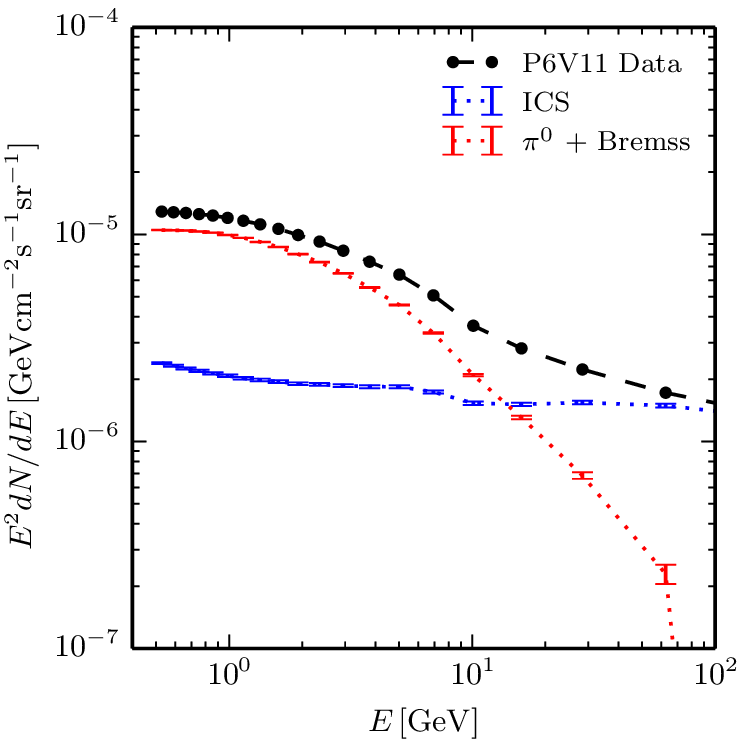}
        \includegraphics[width=0.4\linewidth]{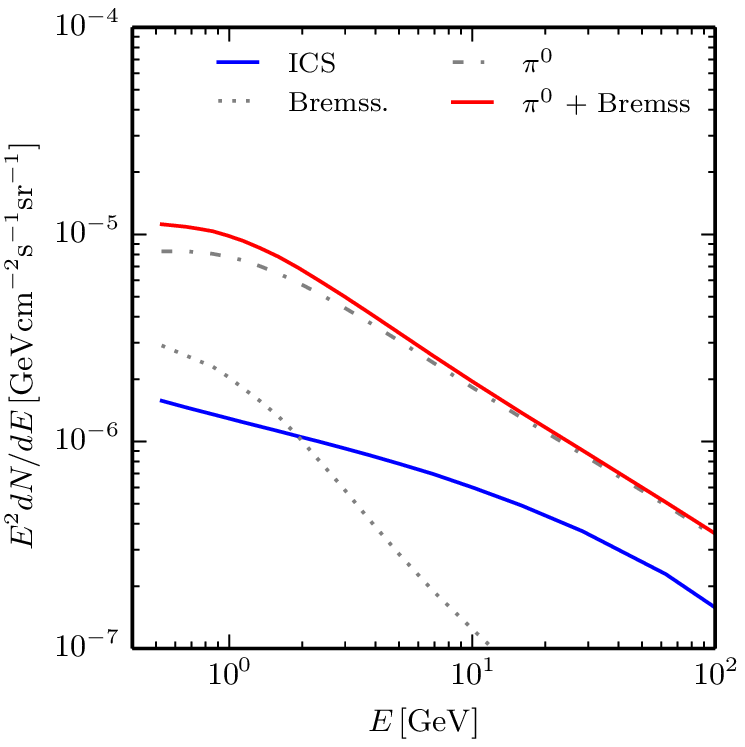}
    \end{center}
    \caption{\textit{Left panel:} ICS and $\pi^0$+Bremss contributions to the 
     \PV background model.  
    \textit{Right panel:} Spectra predicted by a typical Galactic diffuse model for 
    ICS, $\pi^0$ and bremsstrahlung emission. Fluxes in the
    $40^{\circ} \times 40^{\circ}$ ROI, $|b|>2^{\circ}$.}
    \label{fig:P6V11}
\end{figure*}
One of the most challenging results for indirect dark matter searches in recent years
is the discovery of an excess emission  
in the gamma-ray flux from the center of our Galaxy. 
The first indications of such an excess date back to 2009~\citep{Goodenough:2009gk,2009arXiv0912.3828V}. 
Since then, several analyses of gamma-ray data from the Large Area Telescope
 aboard the \Fermi~satellite~\citep{1999APh....11..277G}, hereafter \Fermi-LAT, claimed the existence of the excess above
 the standard astrophysical background at GeV energies~\citep{Goodenough:2009gk, Hooper:2010mq, Boyarsky:2010dr, Hooper:2011ti,
Abazajian:2012pn, Macias:2013vya, Abazajian:2014fta, Daylan:2014rsa,
Zhou:2014lva}. The excess emission
 results from analyses of both the inner few degrees of the Galaxy~\citep{Abazajian:2012pn, Gordon:2013vta,
  Macias:2013vya,
Abazajian:2014fta, Daylan:2014rsa, Zhou:2014lva} and  
 \emph{higher} latitudes~\citep{Hooper:2013rwa, Huang:2013pda, Daylan:2014rsa}, 
 extending up to tens of degrees.
 Intriguingly, the observed spectral energy distribution and the spatial properties 
 of the \Fermi~GeV excess match the expectation for a signal from dark matter particles annihilating
 in the halo of the Milky Way.
 Nevertheless,  some discussion about astrophysical explanations 
 were put forward, as, for example, about
 the emission from a population of point-like sources below the 
 telescope's detection threshold~\citep{Hooper:2013nhl, Calore:2014oga,
Cholis:2014lta, Petrovic:2014xra}, or violent burst events at the Galactic center
with injection of
leptons and/or protons some kilo-/mega-years
ago~\citep{Carlson:2014cwa, Petrovic:2014uda}. 
 
Regardless of the possible interpretations, all analyses agree
on the fact that an extra-emission over the \emph{standard} astrophysical 
background is present in the inner region of the Galaxy.
We stress here that the Galactic center is one of the most promising 
targets for dark matter searches since there the typically predicted profiles
for the dark matter 
distribution lead to the largest photon flux from dark matter origin.
However, the Galactic center is maybe the most challenging target for dark matter searches: 
our knowledge of the conditions at the Galactic center is indeed very poor
and the astrophysical background (from point sources as well as from diffuse emission processes)
is affected by large uncertainties.

A critical point is to answer the question ``An excess above what?".
The excess emission is defined with respect to specific
  astrophysical foregrounds and backgrounds, like
the Galactic diffuse emission (which originates from the interactions of cosmic rays 
with gas and photons in the Galaxy), point-like and extended sources.
Those components should be modelled independently. Therefore, 
it is crucial to explore different foreground and background models in order
to robustly identify and characterise the excess emission.

We will present here part of the analysis performed in~\cite{Calore:2014xka}, 
where we showed for the first time that the excess is statistically robust
against theoretical model systematics, bracketed by exploring previously neglected uncertainties 
on the Galactic diffuse emission, and that the proper treatment of background modelling 
uncertainties allows more freedom for models fitting the excess~\citep{Calore:2014nla}. 

\section{On the importance of foreground modelling}
The dominant source of background for the Galactic center analysis is 
the emission originating from the interaction of cosmic rays with dust and gas in the
Galaxy. The three main production mechanisms of Galactic diffuse gamma rays are:
the Inverse Compton 
scattering (ICS) of electrons on low-energy ambient photons, the decay of abundantly
produced neutral pions and the bremsstrahlung of 
electrons in the interstellar medium.
Most of previous analyses adopted the same background model
to describe the Galactic diffuse emission, namely the \PV background model, provided by
the \Fermi-LAT Collaboration for the \textit{sole} purpose of
point source analysis.\footnote{\texttt{http://fermi.gsfc.nasa.gov/ssc/data/p6v11/access\\/lat/ring\_for\_FSSC\_final4.pdf}}
Using this model for analysis of extended sources introduces systematic effects that
might lead to biased statements 
about the spectrum and morphology of the \Fermi~GeV excess emission.

To visualise this effect, we decomposed the \PV model in the main contributions 
to the Galactic diffuse emission.
 The spectra for ICS, $\pi^0$ and bremsstrahlung
(that we consider as a unique component ``$\pi^0$+Bremss'') are predicted by 
a standard model for cosmic-ray propagation in the Galaxy (see~\cite{Calore:2014xka}
for more details).
We fitted simultaneously the ICS and $\pi^0$+Bremss components to \PV mock-data.
From Figure~\ref{fig:P6V11}, left panel, the reader can see that an extremely 
hard ICS emission at energies $\geq 10$ GeV is an intrinsic property of the \PV
background model. The effect on any analysis that employs it as Galactic diffuse emission
model is to over-subtract the ICS component at high energies, forcing the GeV excess spectrum
to fall-off at $\geq 10$ GeV.

This exercise demonstrates the relevance of modelling \emph{separately}
the different contributions to the Galactic diffuse emission. Indeed, ICS, $\pi^0$ 
and bremsstrahlung possess intrinsically different morphologies because of the
different targets that originate these components: the gas for the $\pi^0$ 
and bremsstrahlung, and the interstellar radiation field for the ICS. Moreover, 
given the different cosmic-ray species responsible of the gamma-ray emission (protons for 
$\pi^0$  and electrons for ICS and bremsstrahlung), also the way in which the morphology
changes with energy is different for the three contributions.

Such arguments strongly motivated the study of 
the variation of the spectral and morphological properties of the excess
due to the modelling of the Galactic diffuse emission.

\section{Home-brew Galactic diffuse emission}
\label{sec:GDE}

\begin{figure*}[t!]
    \begin{center}
        \includegraphics[width=0.8\linewidth]{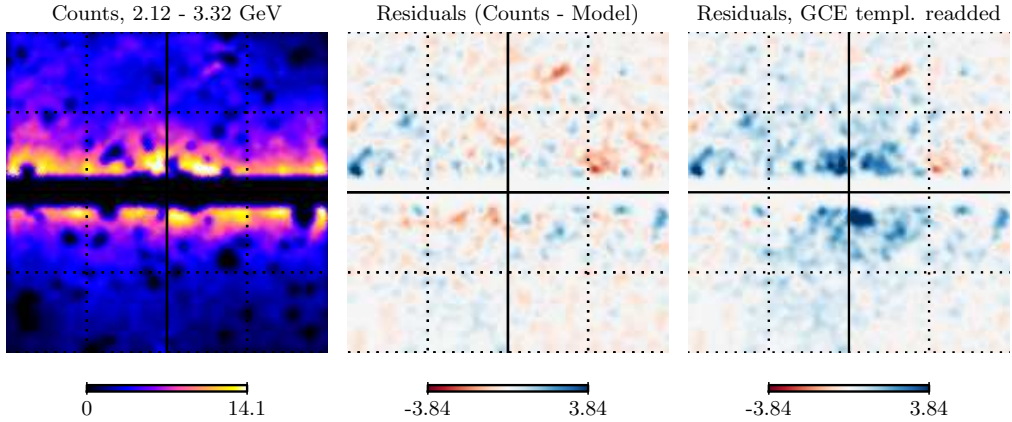}
    \end{center}
    \caption{\emph{Left panels:} Count maps (the $|b|>2^{\circ}$ cut and the
    point-source mask are clearly visible).  \emph{Central panels:} Residuals when subtracting 
    all emission model templates.  \emph{Right panels:} Residuals when 
    re-adding the GeV excess template absorbed emission.}
    \label{fig:residuals}
\end{figure*}

In order to robustly identify the excess despite of large variations in the
foreground emission, we built a set of Galactic diffuse models by varying cosmic-ray
propagation parameters within a given set of assumptions.
We note that the observed emission  results from a line of sight integral and, as
such, it receives contributions from all distances.
In particular, the emission that
  comes from the Galactic center is -- in the models we are adopting --
  relatively subdominant (about 10\%)
and
the Galactic diffuse emission is dominated by local processes.
Therefore, our work should read as the characterisation of
 the uncertainties due to the Galactic gamma-ray emissivity along the line of sight.
We worked in the framework of steady state solutions to the transport equation 
of cosmic-ray propagation in the Galaxy. Homogeneous diffusion, re-acceleration
and convection were considered. We adopted models from the set of~\cite{FermiLAT:2012aa} 
to test variations of the diffusion zone geometry, the source distribution, the spin temperature
and the magnitude cut (for an explanation of cosmic-ray propagation parameters and their 
range of variation see~\cite{Calore:2014xka}).
Additionally, we generated our own Galactic diffuse models
using \texttt{Galprop v54}
(webrun version).  With those models, we  
explored the remaining  uncertainties related
to the diffusion coefficient, re-acceleration, convection, interstellar radiation field,
and Galactic center magnetic field
distributions. In total, we built a set of about 60 models for the Galactic diffuse emission
that test ``extreme'' variations in the parameter space.
We here quote the explored parameter ranges:
\begin{itemize}
    \item geometry of the diffusion zone: $4 \leq z_{D} \leq10$ kpc and $r_{D}$
        = 20 or 30 kpc;

    \item source distributions: SNR, pulsars, OB stars;

    \item diffusion coefficient at 4 GV: $D_{0}  = 2-60 \times 10^{28}$
        $\cm^{2} \s^{-1}$;

    \item Alfv$\acute{\textrm{e}}$n speed: $v_{\rm A} = 0-100$ $\km \s^{-1} $;

    \item gradient of convection velocity:  $dv/dz$ = 0 -- 500 $\km \s^{-1} \kpc^{-1}$;

    \item interstellar radiation field model factors (for optical and infrared emission): 0.5 -- 1.5;

    \item magnetic field parameters: $5 \leq r_{c} \leq 10$ kpc,  $1 \leq z_{c}
        \leq 2$ kpc, and $5.8 \leq B(r=0, z=0) \leq 117$ $\mu$G. 
\end{itemize}

We note that we did not test those models against local cosmic-ray data
and large scale diffuse gamma-ray data (or even
microwave data). 
  
 As already mentioned, we  made a few
simplifying assumptions that we summarise below and that 
 will become relevant for future refined analyses of the GeV excess :
(i) homogeneity and isotropy of cosmic-ray diffusion,
     re-acceleration, and convection;
(ii) radial symmetry of cosmic-ray source distribution in the
        Galactic disk (i.e.~no modelling of the spiral arms), and same 
        source distribution for different cosmic-ray-species sources;
(iii) steady state regime, excluding transient
        phenomena as, for example, burst events.

\section{The data analysis}
In order to analyse gamma rays collected by the \Fermi-LAT from the inner
Galaxy,  we adopt a template-based multi-linear regression technique, 
see, for example, ~\cite{Dobler:2009xz, Su:2010qj}.
The data sample corresponds to 284 weeks of reprocessed \Fermi-LAT data (from
4 August 2008 on) in the energy range 300 MeV -- 500 GeV. 
The Region-Of-Interest (ROI), i.e.~the \emph{inner
Galaxy}, is defined as
\begin{equation}
    |\ell|\leq 20^\circ \quad \text{and} \quad 2^\circ\leq |b| \leq 20^\circ\;,
    \label{eqn:ROI}
\end{equation}
The choice of the latitude cut is such to avoid the large contamination of 
point sources in the innermost few degrees, where the source confusion
is very high.
We prepare the data according to standard prescriptions provided by
the \Fermi-Science-Support-Center.
We binned the data on an \texttt{healpix} grid with resolution
parameter $n_{\rm size}=256$, for each energy bin (24 in total) defined in a way such
to guarantee good statistics also at the highest energies.

We compare the data maps with the model maps, obtained by the superposition of
the different templates adopted in the analysis (see below).
The best-fit normalisation of each model template is derived through a  
maximum likelihood method, based on the Poisson likelihood function 
(cf.~Eq.~(2.3) in~\cite{Calore:2014xka}).

The spatial model templates adopted in the analysis are:
\begin{itemize}
\item Point-like sources template as derived from the 2FGL~\cite{Collaboration:2011bm}, 
with fixed spectra and flux normalisations.
\item \Fermi~bubbles modelled by a uniform-brightness spatial template with bubbles'
 edges as in ~\cite{Su:2010qj}.
\item Isotropic gamma-ray diffuse background with uniform-brightness emission 
template.
\item Galactic diffuse emission ICS and $\pi_0$+Bremss \textit{independent}
templates as modelled from Sec.~\ref{sec:GDE}.
\item GeV excess template whose volume emissivity is parametrised by
 the spherically symmetric generalized NFW profile, 
\begin{equation}
    \rho(r) = \rho_s \frac{(r/r_s)^{-\gamma}}{(1 + r/r_s)^{3 - \gamma}} \,,
    \label{eq:NFWgen}
\end{equation}
squared, and with (best-fit) spectral index $\gamma = 1.2$. This choice is clearly motivated 
by the dark matter annihilation interpretation of
the GeV excess, although we tested a large range of variation for the profile parameters. 
\end{itemize}
The fitted spectra of the \Fermi~bubbles and isotropic diffuse background templates
are constrained to vary within the measured spectra 
from~\cite{FranckowiakBubbl} and~\cite{AckermannEGB}, respectively.

In the analysis, we introduced the following technical improvements:
a non-logarithmic energy binning such to counterbalance the reduced
photon statistics above 10 GeV, a weighted adaptive masking of point sources, and 
the full treatment of the \Fermi-LAT point spread function.

\section{Selection of main results}
\begin{figure}[t!]
    \begin{center}
        \includegraphics[width=1.0\linewidth]{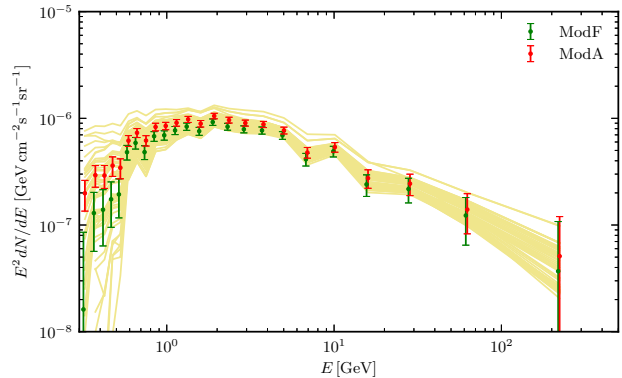}
    \end{center}
    \caption{Spectral energy distribution of the GeV excess template,
    for a generalised NFW profile with an inner slope $\gamma=1.2$. The \emph{yellow band}
    corresponds to all of the 60 GDE models.  Two models are highlighted: 
    the model
    that provides the best fit to the data (model F, \emph{green
    points}) and a reference model (\emph{red points}).}
    \label{fig:spectrum60models}
\end{figure}

\begin{figure*}[t!]
    \begin{center}
        \includegraphics[width=0.75\linewidth]{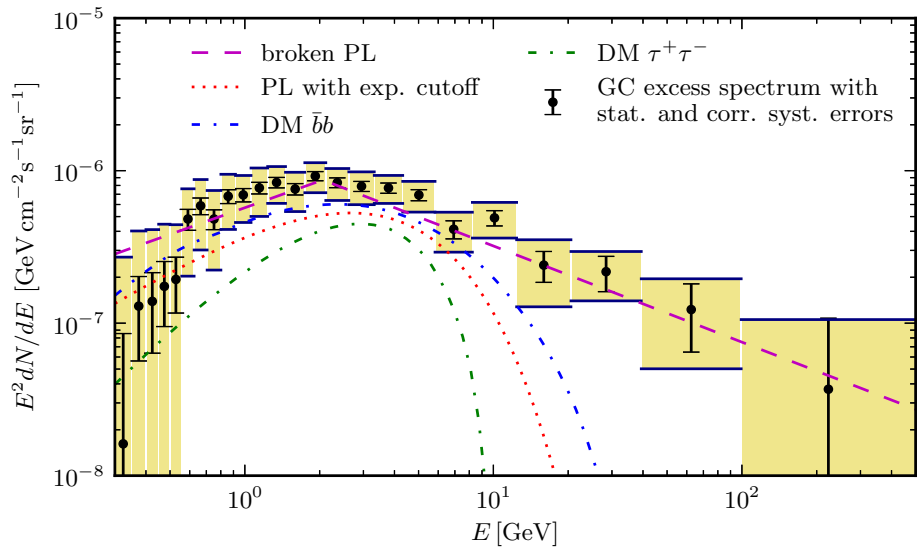}
    \end{center}
    \caption{GeV excess emission spectrum, together with statistical and
    systematical errors, for model F (i.e.~best-fit model). Several spectral
    models  have been fitted to the data. All the spectra (except the $\tau^+\tau^-$) 
    provide a quite reasonable fit to the data. This is due to the
   correlation of the systematic errors (see text).}
    \label{fig:spectrumSysStat_fits}
\end{figure*}

\begin{table*}[t!]
    \small
    \begin{center}
        \begin{tabular}{cccc}
            \hline
            \hline
            Spectrum & Parameters & $\chi^2$/dof & $p$-value \\
            \hline
            broken PL &  $\alpha_1 =  1.42_{-0.31}^{+0.22} $,  $\alpha_2 =  2.63_{-0.095}^{+0.13} $,  $E_{\rm break} =  2.06_{-0.17}^{+0.23} {\rm\ GeV}$ & 1.06 & 0.39 \\[3pt]
            DM $\chi\chi\to\bar{b}b$ &  $\langle\sigma v\rangle =  1.76_{-0.27}^{+0.28} \times 10^{-26}{\rm\ cm^3\, s^{-1}}$,  $m_\chi =  49_{-5.4}^{+6.4} {\rm\ GeV}$ & 1.08 & 0.36 \\[3pt]
            DM $\chi\chi\to\bar{c}c$ &  $\langle\sigma v\rangle =  1.25_{-0.18}^{+0.2} \times 10^{-26}{\rm\ cm^3\, s^{-1}}$,  $m_\chi =  38.2_{-3.9}^{+4.6} {\rm\ GeV}$ & 1.07 & 0.37 \\[3pt]
            PL with exp. cutoff &  $E_{\rm cut} =  2.53_{-0.77}^{+1.1} {\rm\ GeV}$,  $\alpha =  0.945_{-0.5}^{+0.36} $ & 1.37 & 0.12 \\[3pt]
            DM $\chi\chi\to\tau^+\tau^-$ &  $\langle\sigma v\rangle =  0.337_{-0.048}^{+0.047} \times 10^{-26}{\rm\ cm^3\, s^{-1}}$,  $m_\chi =  9.96_{-0.91}^{+1.1} {\rm\ GeV}$ & 1.52 & 0.06 \\
            \hline
            \hline
        \end{tabular}
    \end{center}
    \caption{Spectral fits to the GeV excess spectrum, with $\pm1\sigma$ errors.
    We show best-ft parameters, 
    reduced $\chi^2$, and  corresponding $p$-value.}
    \label{tab:fitResults}
\end{table*}

In this section, we present a selection of the results of the analysis,
 and we refer the reader to~\cite{Calore:2014xka} for a thorough explanation of our findings.
 Figure~\ref{fig:residuals} represents
 the residual (i.e.~data - model counts) emission obtained when subtracting from the raw data the 
 emission associated with the model templates (central panel). The residuals are at the level of 20\% in the whole
 ROI, but, when the GeV excess template associated to the model is re-added (right panel), the residuals in the central
 region of the ROI increase significantly, attesting the presence of the excess, which is, 
 after the other components are subtracted, the most pronounced large-scale excess in our ROI. 
 
Figure~\ref{fig:spectrum60models} represents the spectral energy distribution 
of the excess emission, i.e.~the emission absorbed by the GeV excess template
during the fitting procedure.
The yellow band results from 
\emph{all} the adopted Galactic diffuse models. Such a band   
brackets the uncertainty due to the theoretical modelling of the Galactic diffuse emission
and affecting the extraction of the GeV excess spectrum.  
The GeV excess emission is found to be remarkably stable against the tested variations 
of the Galactic foreground. The typical GeV excess spectrum shows a rising below 1 GeV
(with a spectral index harder than $\sim2$ for all Galactic diffuse models) and features a  peak at
energies around 1--3 GeV.  Despite previous analyses, at higher energies, the spectrum is 
described by a power-law with slope $\sim$ -2.6. The coloured data points indicate
the spectrum (with statistical errors) that corresponds to the best-fit Galactic diffuse model (model F) and 
another exemplary model discussed in~\cite{Calore:2014xka} (model A).

The envelope of the yellow lines corresponds to the \emph{theoretical model uncertainty}, which is due
to the variation induced by the Galactic diffuse modelling. Such uncertainty is, at all energies, larger than the
statistical errors, indicating the importance of the proper treatment of background model systematics.

As it can be already deduced from the residual plots, the Galactic diffuse models 
tested in the present analysis do not describe the data at the statistical level, but, still, 
they show large residuals in the ROI.
Indeed, although the reduced $\chi^2$ for the best-fit Galactic diffuse model (model F) in the energy range from 500 MeV to 3.31
GeV is close to one ($\simeq 1.10$) because of the large number of free parameters, the corresponding
 $p$-value is ridiculously small, $\simeq10^{-300}$. 

On the base of this argument, it is important to find an alternative way of assessing the systematics uncertainties 
affecting the excess. In~\cite{Calore:2014xka}, we relied on an empirical method to derive model systematics due to how well
the different Galactic diffuse models describe the data along the disk, away from the Galactic center.
The derivation and definition of the \emph{empirical model systematics} were presented during this  
conference in a complementary talk, 
``Robust Identification of the GeV Galactic Center Excess at Higher Latitudes".\footnote{C.~Weniger et al., proceedings RICAP-14 (to appear soon).}
Quantifying the background empirical model systematics turned out to be crucial for making statistics based claims
on the possible interpretations of the excess.

\section{Interpretations}
As explained in Sec.~\ref{sec:intro}, several interpretations have been proposed, 
ranging from purely astrophysical to dark matter explanations.
As a first constraint, the predicted model spectrum must provide a good fit to the 
GeV excess spectrum. We performed parametric fits to the GeV excess observed spectrum 
 fully taking into account the systematic uncertainties.

To do so, we made
use of a $\chi^2$ function with a non-diagonal covariance matrix, which models
 the correlated empirical model systematics.
 The $\chi^2$ function used writes as:
\begin{equation}
    \chi^2 = \sum_{ij} 
    \left(\frac{d\bar N}{dE_i}(\vec{ \theta})-\frac{dN}{dE_i} \right) 
    \Sigma^{-1}_{ij}
    \left(\frac{d\bar N}{dE_j}(\vec{ \theta})-\frac{dN}{dE_j} \right) \;,
\end{equation}
with $\Sigma^{-1}_{ij}$ the covariance matrix. 
The
  covariance contains model uncertainties that were derived from the size of
  typical residuals along the Galactic disk.  They amount to variations in the
  excess template that are similar to the ones shown in Figure~\ref{fig:spectrum60models}, and are
  illustrated in Figure~\ref{fig:spectrumSysStat_fits} by the yellow boxes.  For details we refer the
  reader to~\cite{Calore:2014xka}.

We tested several spectra that are related to the GeV excess viable interpretations.
Table~\ref{tab:fitResults} summarises our findings. In particular, parametric fits
 with correlated errors show equal preference for a broken power-law spectrum 
 and for the spectrum from dark matter annihilation into b-quarks. 
 Remarkably, the p-value for a spectrum due to  dark matter annihilation into $\tau$-leptons
 is higher the 0.05. The reason for which dark matter annihilation spectra provide 
 good fit to the GeV excess is due to the fact that systematics errors are correlated in 
 energy and can be understood in terms of the covariance matrix (we refer the interested 
 reader to~\cite{Calore:2014nla}).
 
\section{Conclusion}
The analysis of the \Fermi-LAT data performed in~\cite{Calore:2014xka}
confirmed the presence of an excess emission in the \emph{inner Galaxy}
and some of its, previously found, specific properties, such as the 2--3 GeV peaked spectral energy distribution, 
the extension to high latitudes and the compatibility with a spherically symmetric spatial distribution.
Those properties were demonstrated to be  remarkably stable against theoretical model systematics, due 
to the variations in the modelling of the Galactic diffuse emission.

We assess empirical model systematics from a scan of the gamma-ray 
flux along the disk and we used those uncertainties as a proxy for the systematics 
affecting the GeV excess at the Galactic center.

Contrary to previous results, we do \emph{not} confirm the
fall-off of the GeV excess spectrum at $E \gtrsim10$ GeV, but 
we do find a high energy tail of the spectrum extending up to 
100 GeV.
However, when we properly treat model systematics and
 include them in the spectral fits as 
correlated errors, we demonstrated that it is possible to equally well fit 
the excess spectrum with both a broken power-law and a
gamma-ray spectrum typically expected from  dark matter particles
 annihilation into $\bar{b}b$ final states. This implies a large, previously neglected,
 freedom for 
 models fitting the GeV excess.
 
\bigskip % extra skip inserted
\begin{acknowledgments}
We are grateful to the organisation of the  V \Fermi~Symposium for
the possibility to present this work during the Conference. 
F.C. acknowledges support from the
European Research Council through the ERC starting grant WIMPs Kairos, P.I. G.
Bertone.
This work has been supported by the US Department of Energy (FERMILAB-CONF-15-037-A)
The authors wish to thank JACoW for their guidance in preparing
this template.

\end{acknowledgments}

\bigskip % extra skip inserted
\bibliography{GCdiff.bib}

\end{document}